%% file: paper.tex
\documentclass{juliacon}
\usepackage[LGR, T1]{fontenc}
\usepackage[utf8]{inputenc} 
\usepackage{amssymb} % this
\usepackage{amsmath}
\usepackage{mathrsfs} % and this together. damn
\makeatletter
\renewcommand\subsection{\@startsection{subsection}{2}{\z@}%
{-3.25ex\@plus -1ex \@minus -.2ex}%
{1.5ex \@plus .2ex}%
{\normalfont\bfseries}}
\makeatother
\input{orcid.tex}

\usepackage{newunicodechar}
\newunicodechar{ä}{\"a}
\newunicodechar{ü}{\"u}
\newunicodechar{ö}{\"o}
\newunicodechar{Ä}{\"A}
\newunicodechar{Ü}{\"U}
\newunicodechar{Ö}{\"O}
\newunicodechar{ß}{\ss}

\DeclareSymbolFont{mathscrLC}{OT1}{pzc}{m}{n} % Chancery for lowercase
\DeclareMathSymbol{\scrd}{0}{mathscrLC}{`d}
\newunicodechar{∂}{\ensuremath{\scrd}}
\newunicodechar{σ}{\ensuremath{\sigma}} %(U+03C3)
\newunicodechar{α}{\ensuremath{\alpha}}
\newunicodechar{β}{\ensuremath{\beta}}
\newunicodechar{θ}{\ensuremath{\theta}}
\newunicodechar{μ}{\ensuremath{\mu}}
\newunicodechar{ν}{\ensuremath{\nu}}
\newunicodechar{⊙}{\ensuremath{\odot}}
\newunicodechar{κ}{\ensuremath{\kappa}}
\newunicodechar{η}{\ensuremath{\eta}}
\newunicodechar{∫}{\ensuremath{\int}}
\newunicodechar{ℓ}{\ensuremath{\ell}}
\newunicodechar{√}{\ensuremath{\sqrt{}}}
\newunicodechar{⊗}{\ensuremath{\otimes}}
\newunicodechar{π}{\ensuremath{\pi}}
\newunicodechar{ℝ}{\ensuremath{\mathbb{R}}}
\usepackage{booktabs}
\usepackage{multirow}

\usepackage{inconsolata}
\usepackage[normalem]{ulem} %\sout
\usepackage{centernot} %\centernot
\usepackage[colorinlistoftodos]{todonotes}
\usepackage{hyperref}
\include{moritzdefs}
\begin{document}

\input{header}

%\title{Applied Measure Theory for Probabilistic Modeling}
%\author[1,*]{Chad Scherrer\orcidA{}}
%\affil[1]{Informative Prior}

%\author[2]{Moritz Schauer\orcidB{}}
%\affil[2]{Chalmers and Gothenburg University, Sweden}
\affil[*]{Corresponding author: \url{chad.scherrer@gmail.com}}

%\keywords{Julia, measure theory, probability, statistics}

\maketitle

\begin{abstract}
Probabilistic programming and statistical computing are vibrant areas in the development of the Julia programming language, but the underlying infrastructure dramatically predates recent developments. The goal of \texttt{MeasureTheory.jl} is to provide Julia with the right vocabulary and tools for these tasks. 

In the package we introduce a well-chosen set of notions from the foundations of probability together with powerful combinators and transforms,
giving a gentle introduction to the concepts in this article.

The task is foremost achieved by recognizing \href{https://en.wikipedia.org/wiki/Measure_(mathematics)}{\emph{measure}} as the central object. This enables us to develop a proper concept of \href{https://en.wikipedia.org/wiki/Probability_density_function}{\emph{densities}} as objects relating measures with each others. As densities provide local perspective on measures, they are the key to efficient implementations.

The need to preserve this computationally so important locality leads to the new notion of \emph{locally-dominated measure}, solving the so-called ``base measure problem'' and making work with densities and distributions in Julia easier and more flexible.

%Because it will be new to many who can benefit from this package, this article serves a dual role of a gentle introduction to the concepts of measure theory (the field), as well as an overview of the core capabilities of \verb|MeasureTheory.jl| (the package).
\end{abstract}

%%%%%%%%%%%%%%%%%%%%%%%%%%%%%%%%%%%%%%%%%%%%%%%%%%%%%%%%%%%%%%%%%%%%%%%%%%%%%%%%%%%%%%%%%%%%%%%%
\section{Why measures?}

\emph{Distributions are an insufficient abstraction for probabilistic modeling.}

Let's first consider Bayesian modeling. In the posterior density of the
parameter $\theta$ given the observation or data $x$,
\[
p(\theta \mid x) = \frac{p(\theta)\ p(x \mid \theta)}{p(x)}\ ,
\]
the denominator is called the \href{https://en.wikipedia.org/wiki/Marginal_likelihood}{\emph{model evidence}}. Computing it efficiently is often difficult or unfeasible.

Some sampling algorithms circumvent this problem and only require knowing the density up to a constant factor.
We then can work with the ``unnormalized posterior density'' instead, but we want to distinguish it from a proper probability density, because that difference determines which algorithms are available.

In developments like this using distributions, it's common to begin in terms of a distribution, but then carry around the unnormalized posterior as a function. Fortunately, the meaning of this function is usually clear from context. But structurally, this representation is now divorced from its meaning. It's no longer a Distribution object, so the tools from the original library can no longer help. This is a perspective we will take often.

Thus, starting with elements of our class of interest, a simple and common operation has led us to something outside of this class. This is analogous to the way polynomials over the reals lead to the complex numbers. If ``the shortest path between two truths in the real domain passes through the complex domain'' (Hadamard), then the same argument can be made for measures as connection between distributions.

As a second example, people working in Bayesian modeling sometimes use \href{https://en.wikipedia.org/wiki/Prior_probability#Improper_priors}{\emph{improper priors}}. Whatever one's position on the merits of this, we think it's reasonable to make this approach possible. But an improper prior does not integrate to one, so it's not a distribution.

Also, other computational and statistical inverse problems can often be framed as Bayesian inversion problem with corresponding ``prior'' for example taking the role of a regularization parameter that might be improper, or even flat as in maximum likelihood estimation.

A final and very different concern is the structure of most distribution libraries, in which distributions are classified primarily according to whether they are ``discrete'' or ``continuous''. Such systems often lack facilities to use these in combination; some go so far as to encode the distinction in the type system, making them fundamentally incompatible.

But ``discrete vs continuous'' is a false dichotomy. For example, in three dimensions we can (and often do) work with distributions over points, lines, planes, or the entire space, or over spaces like a simplex or the surface of a  sphere. These can be combined in rich ways, for example as a spike and slab prior for sparse modeling, or as a parameterized subspace for low-rank modeling.

We can address all of these points by extending the system we work with. Instead of distributions, our primary class of interest is \emph{measures}, with distributions as a special case.

Of course, this special case is particularly useful. The point is not to disregard distributions, but to change our focus to one allowing a richer calculus for reasoning. Adapting Hadamard's argument, 
\emph{The shortest path between two distributional truths passes through  non-distributional measures}.

\subsection*{Contributions}

Our work is novel in several ways. \verb|MeasureTheory.jl| has
\begin{enumerate}
    \item[$\bullet\!$]Explicitly represented \emph{base measures} with the same sophistication as the rest of the system, in particular more than just ``discrete or continuous'';
    \item[$\bullet\!$]A \emph{local} approach for determining \href{https://en.wikipedia.org/wiki/Absolute_continuity}{\emph{absolute continuity}}, which is usually a global characteristic;
    \item[$\bullet\!$]Multiple parameterizations for a given measure, without a proliferation of constructors; and
    \item[$\bullet\!$]Normalization and support constraints held separately from the data-dependent computation, allowing for greater efficiency.
\end{enumerate}

Some parts of our approach can been seen in existing systems, though these are still far from universal:
\begin{itemize}
    \item[$\bullet\!$] A rich set of \emph{combinators} for building new measures from existing ones;
    \item[$\bullet\!$] Flexible type constraints, for example allowing measures with symbolic parameters;
    \item[$\bullet\!$] Light-weight measure construction, replacing a common assumption that once constructed, a measure will be used many times. This is especially important for probabilistic programming applications.
\end{itemize}

%%%%%%%%%%%%%%%%%%%%%%%%%%%%%%%%%%%%%%%%%%%%%%%%%%%%%%%%%%%%%%%%%%%%%%%%%%%%%%%%%%%%%%%%%%%%%%%%
\section{What are measures?}

We'll now describe some foundations to help the reader get a deeper understanding of our approach. In particular, we'll define \emph{measures} and \emph{probability distributions}, relate them to familiar notions such as \emph{volume} and the \emph{probability of an event}.

Throughout this discussion, it's important to keep in mind that for us the measure-theoretic abstractions are only means to an end.% 
\footnote{The stackoverflow discussion \url{https://mathoverflow.net/q/11591} discusses books on measure theory; M.S.~likes \cite{Shiryaev1996} as starting point and uses \cite{Elstrodt2011} and \cite{Bauer_Heinz1982-01-01} as reference. As we anticipate interest in connecting with the actual theory, we give some pointers to more technical details in the footnotes.}
Our primary interest, and the goal of \verb|MeasureTheory|, is to offer support for \emph{applied probabilistic modeling}. 
The package name is owed to the fact, that the word measure alone is too vague.

Given a space $X$ of possible outcomes and a set of subsets of $X$ called \href{https://en.wikipedia.org/wiki/Event_(probability_theory)}{\emph{events}}, a probability distribution assigns each event a non-negative quantity, called the \emph{probability} or \emph{probability mass}.
Likewise, a \emph{measure} assigns each \emph{measurable set} a non-negative quantity, also called a ``mass''.

The space $X$ could be the space $\{1,2,3,4,5,6\}$ suitable to model a six-sided die, or $X$ might be the 3-dimensional Euclidean space $\RR^3$ suitable to model ``volume'', to give two examples.

One further such space that deserves mention is traditionally denoted $\Omega$. This is an abstract space connected to the real-world notion of \emph{probability}, which we denote as $\operatorname{Prob}$.
Computationally, $\Omega$ can be considered to be the set of possible initial states of a random number generator \verb|rng::AbstractRNG|, which is the source of computational randomness.

$\Omega$ has a special role of tying random quantities a program produces and their
distributions together, through the notion of random variables, which are
functions $\bX$, $\bY$ from $\Omega$ taking values in spaces $X$, $Y$ such as
$\{1,2,3,4,5,6\}$ or $\RR^3$.\footnote{Additionally, one requires a random
variable to be a \href{https://en.wikipedia.org/wiki/Measurable_function}{\emph{measurable
function}}, that is, one for which the inverse image of an event $\subset X$ is
also an event of $\Omega$.
This is a similar, but much weaker requirement to the definition of a continuous function (``the inverse image of an open set is also open'').}
The package \verb|Omega.jl| \cite{tavares2019language} uses random variables
derived from $\Omega$ as  core principle. 

From a computational perspective,
a Julia function taking only an \verb|rng::AbstractRNG| argument, making a number of calls to \verb|rand(rng)|, and returning a value $x\in X$ is a random variable $\bX \colon \Omega\to X$.
A simple example is  
\begin{verbatim}
    X(rng=Random.GLOBAL_RNG) = rand(rng, 1:6)
\end{verbatim}
While mathematically $\bX$ is a function from $\Omega$, the argument $\omega$ is often hidden like the function argument \verb|rng| of its computational counterpart---not a coincidence.

Each random variable $\bX$ is tied to its \emph{probability law} $D$, the distribution assigning probability to the events $\{\bX \in A\}$ for each (measurable) set $A$
\[\operatorname{Prob}(\bX \in A) = D(A).
\]
In that sense, classical distribution packages restrict themselves very much to the task of providing a catalogue of useful probability laws, and a simple mechanism to provide a random draw (a random variable) with that law, as function
\verb|rand(rng, D)|.
In the example, 
\[\operatorname{Prob}(\bX \in A) = D(A) =  \dfrac{|A|}{6}\]
where $|A|$ denotes the number of elements of $A \subset \{1,2,\cdots,6\}$.

In \verb|MeasureTheory|, measures have abstract super-type \verb|AbstractMeasure|.

\subsection*{Kolmogorov's axioms}

We now introduce \href{https://en.wikipedia.org/wiki/Probability_axioms}{\emph{Kolmogorov's axioms}}, which describe laws that characterize probability distributions, and, with one exception, also measures. 

Measures and probability distributions are both required to obey the axiom that the (probability) mass of sets obtained as union of disjoint component sets\footnote{Precisely: union of a sequence of disjoint component sets.}, equals the sum of the (probability) masses of the components. So if sets $A$ and $B$ are \emph{not} disjoint, the mass of the union is computed from the mass of the components using \emph{inclusion-exclusion}. For probabilities this is
\[
P(A \cup B) = P(A) + P(B) - P(A \cap B)\ .
\]
Similarly for a measure $\mu$,
\[
\mu(A \cup B) = \mu(A) + \mu(B) - \mu(A \cap B)\ .
\]
%which can also be written as
%\[
%\int\limits_{A \cup B} d\mu = \int\limits_{A} d\mu + %\int\limits_{B} d\mu - \int\limits_{AAp B} d\mu\ ,
%\]

In the case where $A$ and $B$ are disjoint, this reduces to 
\[
P(A \cup B) = P(A) + P(B)\ ,
\]
a property called \emph{additivity}.\footnote{The axioms also require a more general form of this to hold. If $\left\{A_n \mid n\in\NN\right\}$ are \href{https://en.wikipedia.org/wiki/Disjoint_sets}{\emph{pairwise disjoint}} (that is, no two have any common elements), 
then it must be true that
\[
P\left(\bigcup_{n\in\NN} A_n\right)
= \sum_{n\in\NN} P\left(A_n\right)\ ,
\]
and similarly for more general measures, replacing $P$ with $\mu$.
This extension of additivity to the countably infinite case is called \emph{$\sigma$-additivity}.}

The reader has encountered many measures before, whether they were given that name. For example, \href{https://en.wikipedia.org/wiki/Counting_measure}{\emph{counting measure}} gives the number of elements of a set, and the \href{https://en.wikipedia.org/wiki/Lebesgue_measure}{\emph{Lebesgue measure}} gives length, area, volume, etc in Euclidean space (\verb|CountingMeasure| and \verb|LebesgueMeasure| in the package.)

So far our characterizations of measures and probability distributions are functionally identical. Indeed, the one distinguishing feature is the \emph{law of unit measure}, which  only for distributions requires that
\[
P(X) = 1\ ,
\]
and also $\operatorname{Prob}(\Omega) = 1$.
% \sout{where $\Omega$ is custom for the set of all possible outcomes, or alternatively,} 
Thus $\Omega$ can be considered the event ``that anything at all will happen''.

By this axiom, probability mass is a proportion, a quantity between 0 and 1. Only this axiom, which for measures is replaced by weaker axiom $\mu(\varnothing) = 0$, sets the two apart. To reinforce that a distribution is also a measure, we'll sometimes refer to it as a \emph{probability measure}.

This close relationship between measures and \emph{probability} measures is illustrated by the Lebesgue measure $\lambda$ on $X=\RR$. The measure of the full space is $\lambda(\RR) = \infty$. But restricting to the unit interval gives $\lambda([0,1])=1$. This \emph{restricted} measure is a probability distribution---the uniform distribution $\uU([0,1])$ describing the law of Julia function \verb|rand()| giving random numbers in the interval $[0,1]$.

We have not yet stated \emph{for which} sets Kolmogorov's axioms have to hold for something to be called a measure or a probability:

\subsection*{$\sigma$-algebras and how to like them}

%As stated at the start of the previous section, 
A (probability) measure on $X$ need not assign a (probability) mass to every possible subset $A\subset X$,
but only to each \emph{event} for a probability measure, or \emph{measurable set} more generally.
Statements we make about sets should be understood as restricted to this set-of-sets, which can be different from one measure to the next.

Now, Kolmogorov's axioms require a given measure to be defined for $X$, and they also require that this set of measurable sets is closed under the operations which the axioms allow (complements and unions of sets or sequences of sets). This imposes an algebraic structure to the measurable sets, called a \emph{$\sigma$-algebra}, which is exactly that: A set of sets $\subset X$ (or $\subset \Omega$ in particular) closed under complementation and countable unions and intersections.

This raises the question: Why not just use the \href{https://en.wikipedia.org/wiki/Power_set}{\emph{power set}}, the ``set of all sets'', which is, after thinking about it, such a $\sigma$-algebra?

A key observation is that having defined probability mass for a number of relatively simple sets, for example on all intervals,
Kolmogorov's axioms give a recipe to compute the probability mass of more complicated sets, and with those, even more in new applications of the axioms, etc.
One says the $\sigma$-algebra is \emph{generated} by these sets.
Indeed, \verb|Distributions.jl| does not give means to compute probability for arbitrary subsets of floating point numbers,
but gives a way of computing probability mass just of intervals of the form $(-\infty, x]$ through the function \verb|cdf|.

For spaces that are ``small enough'', a meaningful probability can be assigned to each subset of the space. 
But even in moderately sized spaces this can become infeasible: Just representing a single arbitrary subset of the \verb|Float64|-range, the computational abstraction of the real line, would require staggering 2\,306 Petabytes of memory. This is not practical. And for any space containing a continuum such as the real line it becomes not impractical but mathematically impossible to assign a meaningful probability to all subsets. 

But even the mathematical measure theory needs to stay clear of the set of all subsets of uncountable spaces, because it turns out that not even volume can be properly defined for all subsets of the Euclidean space.\footnote{Another thing to keep in mind is that, for a function $f\colon X \to Y$ to be measurable requires that ``the inverse image of a measurable set must be measurable''. So allowing more sets to $B\subset Y$ to be measurable requires either allowing more measurable $A\subset X$, or restricting the set of measurable functions.}

Accordingly, as of this writing, \verb|MeasureTheory| does \emph{not} have first-class $\sigma$-algebras, but rather considers them to be implicit to a given measure. 
Having said that, $\sigma$-algebras do have a role to play in applied measure theory, but we give only a pointer here. 
The practical importance of $\sigma$-algebras here lies therein that they also encode available information. Observe that each event $A$ corresponds to a question, with a probabilistic answer. 
 For example, an event $A = \{\bX \in  [c, d]\}$ corresponds to the question whether the random variable $\bX$ is in the interval $[c, d]$ and there is a probability $\operatorname{Prob}(A)$ that this will be the case. 
 Assume we know whether some of such questions $A$, $B$ are answered (each with yes or no).
 
 Then we also know that the answer to the complement $A^c$ (no or yes), and we always know that $\Omega$ has probability 1 and the corresponding question is answered by `yes'. Likewise, if both $A$ and $B$ can be answered given what we know, we also know the answer to $A \cup B$. 
 
 As the reader might now suspect, such systems of known events form a $\sigma$-algebra. In particular, for each random variable $\bX$ there is a $\sigma$-algebra associated with $\bX$ that encodes what else we know knowing $\bX$ and how we should assign (conditional) probabilities to other events knowing $\bX$.
 
 This is a very relevant question for probabilistic programming tasks and causal inference. But we've already mentioned that \verb|MeasureTheory| does not explicitly represent $\sigma$-algebras. Instead, questions like these can be addressed using \emph{kernels} (Section~\ref{Kernels}), structures which can hold the conditional probabilities given we find a way to assign them in a computationally efficient way.

\subsection*{Densities}

A measure is often described in terms of its \emph{density}. Before getting to a more technical discussion, we hope a physical analogy can help build some intuition. Imagine a wooden board with some knots in it, where we might be interested in the mass of some two-dimensional cut-out shape. This mass depends not only on the shape but also on its location and orientation, in particular the inclusion of knots. In this way we can start with a measure (here Lebesgue measure) and use a \emph{density} (the physical density of the wood) to construct a new measure (the mass of any given cut-out shape).

Of course, we do this all the time with distributions, building a continuous distribution in terms of a \emph{probability density function (pdf)} over Lebesgue measure, or a discrete distribution in terms of a \emph{probability mass function (pmf)}, which is just a density over counting measure. 

Somewhat more formally, a probability density in Euclidean space is a \emph{local} ratio of probability assigned to an infinitesimally small volume/area $\dd x$ around each point $x$, relative to that volume itself
\[
f(x) = \frac{\operatorname{Probability}(\dd x)}{\operatorname{Volume}(\dd x)} \ .
\]

``Volume'' is not a distribution, but is in fact the Lebesgue measure described above. Discrete distributions can be expressed similarly using counting measure. 
There are certainly more events (sets) than outcomes (elements), so a probability density gives a local and parsimonious description of a distribution.

More generally, for measures $\mu$ and $\nu$ (and an \emph{absolute continuity} condition, to be discussed), the density is
\[
f(x) = \frac{\mu(\dd x)}{\nu(\dd x)} = \frac{\dd\mu}{\dd\nu}(x)\ .
\]

As we see, density only makes sense relative to some \emph{base measure}.

Limiting ourselves to distributions makes this awkward to even discuss, but allowing measures as first-class objects means we can make this characterization more explicit, and thus more flexible. What sets \verb|MeasureTheory| apart from most libraries is that we don't sweep this under the rug, but rather address it head-on. 

 In place of Lebesgue or counting measure, any measure we can express can play the role of $\dd\nu$ above. This becomes crucial in high-dimensional spaces, where working with other reference measures such as a product of normal distributions becomes a numerical requirement. (Mathematically, there is no infinite dimensional Lebesgue measure, so numerically, even  high-dimensional Lebesgue measures can be problematic.
\cite{https://doi.org/10.5281/zenodo.3931118} for example allows to express a target density with respect to product of normal distributions using the Boomerang sampler \cite{pmlr-v119-bierkens20a} for this reason.)
 
%%%%%%%%%%%%%%%%%%%%%%%%%%%%%%%%%%%%%%%%%%%%%%%%%%%%%%%%%%%%%%%%%%%%%%
\section{Locally dominated measures}
One cannot expect to express a measure $\mu$ putting positive mass on a set $S$ relative to a second measure $\nu$ on $S$, if nothing is there to compare, that is if $\mu(S) > 0$ but $\nu(S) = 0$. You can't ``make something from nothing''.

Given measures $\mu$ and $\nu$ defined on a common space $X$, we say $\mu$ is \emph{dominated by} $\nu$ 
if $\nu(S)=0$ implies $\mu(S)=0$  (or equivalently, $\mu(S)>0$ implies $\nu(S)>0$) for every measurable $S$.
This is denoted by $\mu \ll \nu$, and is sometimes equivalently read as ``$\mu$ is \emph{absolutely continuous} with respect to $\nu$''.

The Radon-Nikodym Theorem states that $\mu \ll \nu$ if equivalently there is a \emph{density} function $f$, often written $\frac{d\mu}{d\nu}:=f$, with the property that for every $S$,
\[
\mu(S) = \int_S f\  \dd\nu \ .
\]

The concept of absolute continuity is very useful for formal manipulations. However, the global nature of testing $\mu \ll \nu$ would require every support to be represented in a way that allows efficient computation of this relation. In particular, in applications based on Markov chain Monte Carlo, such global information about the measures at hand is not available.

Even with such a capability, this approach would have its problems. It seems likely a user getting an error in response to requesting a density would often respond by restricting measures accordingly until the request can be fulfilled.

Because of this, we instead define $\mu$ to be \emph{locally dominated} by $\nu$ near $x$, written $\mu \ll_x \nu$, if there is some neighborhood $N \ni x$ such that $\nu(N)>0$ (so it's not a degenerate case) and $\mu|_N \ll \nu|_N$ (absolute continuity between the restricted measures). A \emph{local density} can be defined similarly. In \verb|MeasureTheory|, the \verb|density| and \verb|logdensity| functions work in exactly these terms. For the remainder of this paper, ``(log-)density'' will always refer to the local (log-)density, and ``local'' will often be taken as understood.

Density computations are typically done in log space. For example, if we're interested in $\frac{\dd \mu}{\dd \nu}$, we'll typically instead compute \verb|logdensity_rel(μ, ν, x)|.\footnote{Density and log-density are \emph{always} relative; the \texttt{rel} here is to indicate that we want the density relative to a second measure which is explicitly specified.}
Here and below, we'll often write the math in terms of densities and the code in terms of log-densities, despite unfortunate inconsistency between the two.

There are several benefits to working in log-space. Results are much less likely to overflow or underflow, and the many products and exponents become sums and products, respectively. which are much more efficient to compute and to differentiate.

Working locally and in log-space also gives us a convenient \emph{antisymmetry},
\begin{verbatim}
logdensity_rel(μ, ν, x) == -logdensity_rel(ν, μ, x)
\end{verbatim}

In particular, \verb|logdensity_rel(μ, ν, x)| takes special values in some common cases:
\newcommand{\notllx}{\mathrel{\centernot{\ll}\!\!_x}}

% works now
\begin{center}
\begin{tabular}{@{}rrr@{}}\toprule
%& \multicolumn{2}{c}{$w = 8$}\\\cmidrule{2-3} 
& $\nu\ll_x \mu$& \text{(else)} \\\midrule
$\mu\ll_x \nu$& (finite) & \verb|-Inf|  \\
\text{(else)} & \verb|Inf| & \verb|NaN|  \\\bottomrule
\end{tabular}
\end{center}

%%%%%%%%%%%%%%%%%%%%%%%%%%%%%%%%%%%%%%%%%%%%%%%%%%%%%%%%%%%%%%%%%%%%%%%%%%%%%%%%%%%%%%%%%%%%%%%%

\section{Base measures and log-densities}
It's very useful for users to be able to \emph{call} \verb|logdensity_rel| for arbitrary measures. But \emph{defining} things directly in these terms would require a definition for every pair of possible measures, which is of course intractable.

We'll now discuss the mechanics of defining a measure, followed by a description of the end-user perspective.

\subsection*{Defining a measure}

To define a new measure requires a \emph{base measure}, the \emph{log-density} with respect to that base measure, and a \emph{support}. 
A base measure can be defined locally or globally. In the latter case, the local base measure defaults to 
\begin{verbatim}
basemeasure(μ::AbstractMeasure, x) = basemeasure(μ)
\end{verbatim}
This allows users to define a base measure specific to a given neighborhood if they prefer, or to define a global base measure when that suffices. 

For example, for a standard \verb|Normal| measure, we have

\begin{verbatim}
basemeasure(::Normal{()}) = (1/sqrt2π) * Lebesgue(ℝ)

logdensity_def(::Normal{()} , x) = - x^2 / 2 

insupport(::Normal{()}, x) = true
\end{verbatim}

Note the difference from a typical distribution-oriented implementation; the log-density consists only of $-x^2/2$, with the normalization term pushed into the base measure. 

The normalization term is present for any normal distribution, even those with a different mean or variance. So if we have a product of normals, for example in a regression problem, the normalization can be shared across them. For the simple example of $n$ iid standard \verb|Normal|s, this effectively rewrites the log-pdf
\[
    \sum_{j=1}^n \left(-\frac{1}{2}\log 2π -\frac{x_j^2}{2}\right)
    \quad \text{as} \quad
    -\frac{n}{2}\log 2π + \sum_{j=1}^n \left(-\frac{x_j^2}{2}\right)
\]

Further simplification is clearly possible; doing this in a composable way without sacrificing efficiency is the subject of ongoing work.

Just as \verb|logdensity_def(μ,x)| defines the log-density relative to the default base measure, it's often useful to be able to define the log-density between two measures that are given explicitly. For this, we have the three-argument \verb|logdensity_def(μ,ν,x)|. 

Though this three-argument form is available as a primitive, it would be intractable to use it to define log-densities for every pair of measures. Instead, \verb|basemeasure| forms a \emph{forest} (a collection of trees, with $μ \rightarrow \nu$ if \verb|ν=basemeasure(μ)|). The three-argument \verb|logdensity_def| connects trees to form a graph. For any two measures, \verb|MeasureTheory| can traverse this graph at compile time to determine how to compute the log-density.

\subsection*{The end-user perspective}

From our above discussion of moving terms between the log-density and the base measure, the reader may suspect that this can be done arbitrarily, leaving no ``natural'' log-density for a given measure. This is indeed the case; the log-density of a measure is only determined \emph{relative to a particular base measure}.

Our choice of \verb|basemeasure| doesn't reflect any mathematical invariant, only computational convenience. Accordingly, \verb|basemeasure| and \verb|logdensity_def| should be considered implementation details, and not part of the user-facing interface.

The way \verb|MeasureTheory.jl| is set up, repeated application of \verb|basemeasure| will eventually reach a fix point. That is, some measure will eventually be its own base measure. For such a measure, \verb|logdensity_def| is required to return zero. We refer to the fix point reached in this way from a given measure as that measure's \emph{root measure}.

Instead of \verb|logdensity_def|, users should typically call \verb|logdensity_rel(μ,ν,x)| (which computes $\frac{\dd μ}{\dd ν}(x)$), or \verb|logdensityof(μ,x)|, which gives the log-density relative to the root measure. \verb|logdensityof| was originally defined in \cite{DensityInterface.jl}, and provides a convenient common ground for those not accustomed to thinking of densities as being relative. We expect this to be especially useful for new users.

%%%%%%%%%%%%%%%%%%%%%%%%%%%%%%%%%%%%%%%%%%%%%%%%%%%%%%%%%%%%%%%%%%%%%%%%%%%%%%%%%%%%%%%%%%%%%%%%
\section{Parameterized measures}

A common challenge in building a library of distributions is the choice of \emph{parameterizations}. For example, in Stan \cite{Stan},
a \emph{negative binomial} distribution is parameterized by $\alpha$ and $\beta$, where
\[
\operatorname{NegBinomial}(y \mid \alpha, \beta)
=\binom{y+\alpha-1}{\alpha-1}\left(\frac{\beta}{\beta+1}\right)^{\alpha}\left(\frac{1}{\beta+1}\right)^{y}\ .
\]

In the Julia package \verb|Distributions.jl| \cite{Distributions.jl-2019}, this is instead given by
\[
\operatorname{NegBinomial}(y \mid r, p)
= \binom{y + r - 1}{r - 1} p^r (1 - p)^y \ .
\]

These are equivalent (let $r = \alpha$ and $p = \frac{\beta}{\beta + 1}$), and the inconsistency alone gives some evidence that it might be reasonable to prefer one or the other depending on the circumstances. Yet most libraries only allow one or the other (or yet another alternative) as \emph{the} negative binomial distribution. Other parameterizations require entirely different names.

In \verb|MeasureTheory|, our approach avoids this problem. A \emph{parameterized measure} is defined by a struct of the appropriate type with a named tuple \verb|par| field. For example,

%\inputminted{julia}{structNormal.jl}
\begin{verbatim}
struct NegativeBinomial{N,T} <: ParameterizedMeasure{N}
    par :: NamedTuple{N,T}
end
\end{verbatim}

We can then write 
\begin{verbatim}
    NegativeBinomial(r=10, p=0.75)
\end{verbatim}
or
\begin{verbatim}
    NegativeBinomial(α=10, β=3)
\end{verbatim}

Calls to \verb|rand|, \verb|logdensity_def|, etc then delegate to methods according to the appropriate names. We use \verb|KeywordCalls.jl| \cite{KeywordCalls.jl}, so all names are resolved statically at compile time.

%%%%%%%%%%%%%%%%%%%%%%%%%%%%%%%%%%%%%%%%%%%%%%%%%%%%%%%%%%%%%%%%%%%%%%%%%%%%%%%%%%%%%%%%%%%%%%%%
\section{Kernels\label{Kernels}}

A \emph{kernel} is a (measurable\footnote{That is, the function $x \mapsto \kappa(x)(A)$ must be measurable for every fixed $A$.}) function $\kappa$ that returns a measure. Equivalently, it represents a family of measures parameterized by its argument. Writing $\cM(Y)$ for ``measures on $Y$'', we can write this as
\[
\kappa\colon X \to \cM(Y)\ .
\]

A prominent application is a \href{https://en.wikipedia.org/wiki/Conditional_probability_distribution}{\emph{conditional distribution}}. In this case $\kappa$ is further restricted to be a \href{https://en.wikipedia.org/wiki/Markov_kernel}{\emph{Markov kernel}}, 
\[
\kappa\colon X \to \cP(Y)\ ,
\]
where $\cP(Y)$ represents ``probability measures on $Y$''.
If $\bX$ and $\bY$ are random variables defined on $X$ and $Y$, the kernel $\kappa$ defines a probability measure assigning every measurable $B\subset Y$ probability
\[
P(\bY \in B \mid \bX = x) =: (\kappa(x))(B)\ .
\]

In \verb|MeasureTheory|, a kernel is represented as either a Julia function or an \verb|AbstractTransitionKernel| object.
As a common special case, a \verb|ParameterizedTransitionKernel| pairs a measure constructor with a mapping into its parameter space, and makes the functional relationship between argument of $\kappa$ and the returned measure transparent.

A Markov kernel, in particular, corresponds to the distribution of a \emph{parametrized random variable}, a function with random outcomes (or to ``mechanisms'' in causal inference, for example in \cite{https://doi.org/10.5281/zenodo.1005091}). For example, the kernel
\[
\kappa\colon x \mapsto \text{Normal}(x,\sqrt{x})
\]
corresponds to the random function
\begin{verbatim}
    f(rng, x) = x + √x * randn(rng)
\end{verbatim}
and can be expressed in \verb|MeasureTheory| as
\begin{verbatim}
    κ = kernel(Normal) do x
        (μ=x, σ=√x)
    end
\end{verbatim}
or
\begin{verbatim}
    κ = kernel(Normal; μ=identity, σ=sqrt)
\end{verbatim}

The latter formulation decomposes the kernel into separate ``parameter maps'', making relationships between \verb|x| and the parameters of \verb|κ(x)| more explicit.

This could of course also be expressed as a parameterized measure. The difference is that kernels are lighter weight to build and have all the dynamism of functions, while the static nature of parameterized measures can make it easier to express some optimizations.

\section{Pointwise products and likelihoods \label{Likelihoods}}

The \href{https://en.wikipedia.org/wiki/Pointwise_product}{\emph{pointwise product}} in probability often describes the fusion of information, in this case the fusion of information from prior and from the observations via the likelihood, but also shows up in related situation, such as \href{https://en.wikipedia.org/wiki/Sensor_fusion}{\emph{sensor fusion}} in signal processing.

To begin, let's consider the case of probability distributions dominated by Lebesgue measure.
For a parameter $\theta$ and data $x$, suppose we have a \href{https://en.wikipedia.org/wiki/Prior_probability}{\emph{prior density}} $p(\theta)$ and \href{https://en.wikipedia.org/wiki/Likelihood_function}{\emph{likelihood}} $p(x | \theta)$. Then Bayes's Law gives the posterior density
\[
p(\theta \mid x) = \frac{p(\theta)\ p(x \mid \theta)}{p(x)}\ .
\]

In practice, we rarely know the normalization factor $p(x)$, so we often work in terms of the unnormalized posterior density, and write
\[
{\underbrace{p(θ \mid x)}_\text{posterior}}
 \propto 
{\underbrace{p(\theta)}_\text{prior}}\ 
{\underbrace{p(x \mid \theta)}_\text{likelihood}} .
\]
Because the product does not include normalization, in general it does not integrate to one, and so it's not a probability density function. But there are no such problems as a density for a measure. We now consider this more general formulation.

Given a \emph{parameter space} $\Theta$ and \emph{observation space} $X$, a \verb|Likelihood| $\ell$ consists of 
\begin{enumerate}
    \item a kernel $\ell_κ : \Theta \to \cM(X)$,
    \item an observation $\ell_x \in X$, and
    \item a base measure $\ell_\beta \in\cM(X)$.
\end{enumerate}

A likelihood can be treated as a function $\Theta \to \RR_{\ge 0}$, by defining
\[
    \ell(\theta) = \frac{\dd\ell_\kappa(\theta)}{\dd\ell_\beta}(\ell_x)
\]

Of particular interest is the case where the base measure $\ell_\beta$ is
\[
    \ell_\beta = \hat{\theta} \equiv \arg\max_\theta \frac{\dd\ell_\kappa(\theta)}{\dd\rho}(\ell_x)
\]

for any non-singular measure $\rho$; this corresponds to the \href{https://en.wikipedia.org/wiki/Likelihood_function#Relative_likelihood_function}{\emph{relative likelihood}}. In \verb|MeasureTheory|, we allow the base measure to be specified, but take as a default the \emph{root measure} of $\ell_\kappa(\theta)$.

Note that a likelihood $\ell$ takes a point in $\Theta$ to an evaluated density in a different space (the observation space $X$). In particular, a likelihood is not a measure.

Rather, a given likelihood acts on measures, taking a ``prior'' measure $\mu$ to ``unnormalized posterior'' measure $\mu \odot \ell$. Relative to another measure $\alpha$, this has density

\[
    {\underbrace{\frac{\dd(\mu\odot\ell)}{\dd\alpha}(\theta)}_\text{unnormalized posterior}}
    \equiv
    {\underbrace{\frac{\dd\mu}{\dd\alpha}(\theta)}_{\text{prior}}}\ {\underbrace{\vphantom{\frac{\dd(\mu\odot\ell)}{\dd\alpha}(\theta)}\ell(\theta)}_{\text{likelihood}}}
\]

Treating likelihoods and pointwise products explicitly in this way gives an easy route to many performance optimizations. For example, the likelihood of a linear model can be computed very efficiently in terms of basic linear algebra, and in cases of conjugacy the pointwise product simplifies to a measure that can be expressed in closed form.

%%%%%%%%%%%%%%%%%%%%%%%%%%%%%%%%%%%%%%%%%%%%%%%%%%%%%%%%%%%%%%%%%%%%%%%%%%%%%%%%%%%%%%%%%%%%%%%%
\section{Product and power measures}

Given measures $\mu$ on $X$ and $\nu$ on $Y$, the \href{https://en.wikipedia.org/wiki/Product_measure}{\emph{(independent) product measure}} $\mu \otimes \nu$ is a measure on $X\times Y$. Just as Lebesgue measure on $\RR$ is \emph{generated} by defining its value on intervals, the product measure is generated by defining
\[
(\mu \otimes \nu)(A\times B) = \mu(A)\ \nu(B)\ ,
\]
for any measurable $A\subset X$ and $B\subset Y$. \verb|MeasureTheory| code uses exactly this notation, \verb|μ ⊗ ν|. This can be extended recursively to products of any finite number of measures.

If $\mu$ is defined in terms of a base measure $\alpha$ and likewise $\nu$ over $\beta$, the product measure has base measure $\alpha \otimes \beta$ and density
\[
\frac{\dd(\mu \otimes \nu)}{\dd(\alpha \otimes \beta)} = \frac{\dd\mu}{\dd\alpha}\ \frac{\dd\nu}{\dd\beta}\ .
\]

A special case of this is when $\mu = \nu$. In \verb|MeasureTheory| we refer to this as a \emph{power measure}, written \verb|μ^2| or \verb|μ^n| for higher dimensions. More generally, the second argument can be a \verb|Tuple|, so for example \verb|μ^(2,3)| extends $\mu$ as a product measure over $2\times 3$ matrices.

A second special case is when we have a collection of values together with a kernel. For this we use the \verb|For| combinator. So in a regression model, we might express the response as coming from
\begin{verbatim}
    For(1:n) do j  
        Normal(β * x[j], σ)
    end
\end{verbatim}
This is exactly the product measure
\[
\bigotimes_{j=1}^n \text{Normal}(\beta x_j,\ \sigma)\ .
\]

More generally, suppose we have a measure $\mu$ on a space $X$ and a kernel $\kappa\colon X \to \cM(Y)$. Then we can define a measure on $X \times Y$ as follows.

For a given $x \in X$ with $\mu \ll_x \alpha$, let $\nu_x = \kappa(x)$ have base measure $\beta_x$. Then similarly to the independent product, we have
\[
\frac{\dd(\mu \otimes \nu_x)}{\dd(\alpha \otimes \beta_x)} = \frac{\dd\mu}{\dd\alpha}\ \frac{\dd\nu_x}{\dd\beta_x}\ .
\]

%%%%%%%%%%%%%%%%%%%%%%%%%%%%%%%%%%%%%%%%%%%%%%%%%%%%%%%%%%%%%%%%%%%%%%%%%%%%%%%%%%%%%%%%%%%%%%%%
\section{Superposition and mixtures \label{superposition}}

A \emph{superposition} is the measure-theoretic analog of a mixture model. For measures $\mu$ and $\nu$ with $f=\frac{\dd\mu}{\dd\alpha}$ and $g=\frac{\dd\nu}{\dd\beta}$, the superposition, written $\mu + \nu$, is a measure with base measure $\alpha + \beta$ and density

\[
\begin{aligned}\frac{\dd(\mu+\nu)}{\dd(\alpha+\beta)} & =\frac{f\,\dd\alpha+g\,\dd\beta}{\dd\alpha+\dd\beta}\\
 & =\frac{f\,\dd\alpha}{\dd\alpha+\dd\beta}+\frac{g\,\dd\beta}{\dd\alpha+\dd\beta}\\
 & =\frac{f}{1+\frac{\dd\beta}{\dd\alpha}}+\frac{g}{\frac{\dd\alpha}{\dd\beta}+1}\\
 & =\frac{f}{1+\left(\frac{\dd\alpha}{\dd\beta}\right)^{-1}}+\frac{g}{\frac{\dd\alpha}{\dd\beta}+1}\ .
\end{aligned}
\]

Using the inverse in this final line allows $\frac{\dd\alpha}{\dd\beta}$ to computed once and then re-used. Also note that in the special case where $\alpha=\beta$, this reduces to $\frac{f+g}{2}$.

An important special case of superposition occurs when $\mu$ and $\nu$ are finite measures on some space $\Omega$, and $\mu(\Omega) + \nu(\Omega) = 1$. This is equivalent to a convex combination of probability densities, called a \href{https://en.wikipedia.org/wiki/Mixture_distribution}{\emph{mixture}}.

Conveniently, a product of measures distributes over superposition. That is, if $\alpha$ and $\beta$ are measures on a common space $X$ and $\gamma$ and $\delta$ are measures on $Y$, then
\[
\alpha \otimes (\gamma + \delta) = \alpha \otimes \gamma + \alpha \otimes \delta\ ,
\]
and
\[
(\alpha + \beta) \otimes \gamma = \alpha \otimes \gamma + \beta \otimes \gamma\ .
\]

A very common special case of superposition is a \href{https://en.wikipedia.org/wiki/Spike-and-slab_regression}{\emph{spike and slab prior}}, a mixture of a \href{https://en.wikipedia.org/wiki/Dirac_measure}{\emph{Dirac measure}} (a point mass) with a continuous measure. 
This is useful for sparse Bayesian modeling, as implemented in the \emph{sparse ZigZag sampler}, described in \cite{bierkens2021sticky} and implemented in \cite{https://doi.org/10.5281/zenodo.3931118} using \verb|MeasureTheory|.

%%%%%%%%%%%%%%%%%%%%%%%%%%%%%%%%%%%%%%%%%%%%%%%%%%%%%%%%%%%%%%%%%%%%%%%%%%%%%%%%%%%%%%%%%%%%%%%%
\section{Density decomposition}

Especially for probability measures, it's common for the log-density with respect to Lebesgue or counting measure to have several types of terms. Given an observation $x$ and any relevant parameters, there are typically 
\begin{itemize}
    \item[$\bullet\!$] Terms that are \emph{data-dependent}, each involving some nontrivial function of $x$ (and possibly also of the parameters),
    \item[$\bullet\!$] Terms that are \emph{parameter-dependent}, and
    \item[$\bullet\!$] Terms that are \emph{constant}.
\end{itemize}

In addition, it's common to have an ``argument check'' to be sure $x$ is in the support of the distribution.

Depending on the application, we can often ignore some of these. For example, we may know $x$ is in the support by construction, or from a previous check. In these cases, any use of resources to check arguments is wasteful.

There are many cases where it's important for performance to have constant and parameter-dependent terms separated from data-dependent ones. Suppose $\mu$ is a measure with log-density of the form
\[
\log \frac{\dd\mu}{\dd\nu}(x) = \ell(x; \theta) = f(x,\theta) + g(\theta) + C\ .
\]

If we instead observe an iid product of $N$ observations, the log-density is
\[
\begin{aligned}
\sum_{j=1}^N\ell(x_j; \theta) 
    &= \sum_{j=1}^N \left[ f(x_j,\theta) + g(\theta) + C\right] \\
    &= N [g(\theta) + C] + \sum_{j=1}^N  f(x_j,\theta)\ .
\end{aligned}
\]

This final form can be much more efficient, because it reduces $N$ repeated computations of $g(\theta) + C$ to one. 

In some cases it's important to avoid computing $g(\theta) + C$ at all. For correlation matrices, it's common to use the LKJ prior \cite{LEWANDOWSKI20091989}. 

Rather than work with correlation matrices directly, it's convenient to work in terms of the Cholesky decomposition. For this purpose, \verb|MeasureTheory| includes an \verb|LKJCholesky| measure. This is typically used as a prior with fixed parameters $k$ and $\eta$, which give the dimensionality and ``concentration'' of the measure. The relative cost of normalization (which is irrelevant for MCMC) can be computed as

\begin{verbatim}
function relative_normcost(k, η)
    μ = LKJCholesky(k, η)
    L = rand(μ).L
    f_cost = @belapsed logdensity($μ, $L)
    g_plus_C_cost = @belapsed Dists.lkj_logc0($k, $η)
    return g_plus_C_cost / (g_plus_C_cost + f_cost)
end
\end{verbatim}

So for example, in 10 dimensions with $\eta=2.0$ this gives
\begin{verbatim}
    julia> relative_normcost(10, 2.0)
    0.76428528899935    
\end{verbatim}

That is, 76\% of the time is spent in normalization. If our application doesn't need it, three-fourths of the computation time is simply wasted.

For these reasons, we break the representation of the log-density into several pieces (now additive terms in log-space):
\begin{enumerate}
    \item[$\bullet\!$]\emph{Constant} terms,  $-\frac{1}{2}\log 2\pi$\ .
    \item[$\bullet\!$]\emph{Parameter-dependent} terms,  $-\log \sigma$\ .
    \item[$\bullet\!$]\emph{Data-dependent} terms, $-\frac{1}{2}\left(\frac{x - \mu}{\sigma}\right)^2$\ .
\end{enumerate}

The two-argument \verb|logdensity| then computes only the data-dependent terms, with the constant and parameter-dependent terms pushed to the base measure. This makes it easy to defer computation of these terms until they are required.

%%%%%%%%%%%%%%%%%%%%%%%%%%%%%%%%%%%%%%%%%%%%%%%%%%%%%%%%%%%%%%%%%%%%%%%%%%%%%%%%%%%%%%%%%%%%%%%%
\section{Affine transforms}
\label{affine}

A particularly expressive way to build new measures in terms of existing ones is through a \emph{pushforward}. Given a measure $\mu$ and a function $f$ defined on its support, the \href{https://en.wikipedia.org/wiki/Pushforward_measure}{\emph{pushforward of $\mu$ through $f$}} is a measure $f_*\mu$ defined by
\[
f_*\mu(S) = \mu(f^{-1}(S))\ .
\]

In the following sections, we first discuss in the context of probability measures before the more general case.

\subsection*{Forward parameterization}

Starting with a $k$-dimensional multivariate random variable $z$, an \href{https://en.wikipedia.org/wiki/Affine_transformation}{\emph{affine transform}} is a ``linear transform with a shift''. We can use this to define a new random variable $x$, as
\[
\bX = \sigma \bZ + x_0\ ,
\]

with $\sigma$ and $x_0$ of the appropriate dimensions.\footnote{We would typically use $\mu$ in place of $x_0$, if not for the unfortunate potential confusion with $\mu$ as a name for a measure.} If $\EE[\bZ]=0$, then $x$ has mean
\[
\EE[\bX] = \sigma \EE[\bZ] + x_0 = x_0
\]
and variance matrix
\[
\VV[\bX] = \EE[(\bX-x_0)(\bX-x_0)^t] = \EE[\sigma \bZ \bZ^t \sigma^t] = \sigma \VV[\bZ] \sigma^t\ .
\]

Note that if $\VV[\bZ]=I_k$, we get $\VV[\bX]=\sigma \sigma^t$, so we can arrive at a given positive semidefinite $\VV[\bX]=\Sigma$ by taking $\sigma$ to be its \href{https://en.wikipedia.org/wiki/Cholesky_decomposition}{\emph{lower Cholesky factor}}. Also, in the special case of a one-dimensional Gaussian, this gives the familiar $\VV[\bX] = \sigma^2$.

We call this the \emph{forward parameterization} because it's especially convenient for sampling, sometimes referred to as ``running the model forward''. Unfortunately, the cost of this convenience is a relatively awkward expression for the density. In this case, we start with $x$ and need to solve $z$, finally adjusting according to the determinant of the transformation:
\[
p_\bX(x) = \frac{1}{|\sigma|}\ p_\bZ(z) = \frac{1}{|\sigma|}\ p_\bZ\left(\sigma^{-1} (x - x_0)\right)\ ,
\]

where $|\sigma|$ is the \href{https://en.wikipedia.org/wiki/Determinant}{\emph{determinant}} of the square matrix, and the second equality comes from solving for $z$, which gives $z=\sigma^{-1}(x-x_0)$. More generally (when the transform is not expressed as a matrix), this role is played by the determinant of the \href{https://en.wikipedia.org/wiki/Jacobian_matrix_and_determinant}{\emph{Jacobian}}, $\left|\frac{\dd x}{\dd z}\right|$.

This requires solving a linear system. Even with $\sigma$ being lower-triangular, this involves division operations and the allocation of a temporary vector for storage of $z$.

In many cases, we prefer the density (or log-density, really) to be fast to evaluate. This leads us to a kind of dual approach to the above.

\subsection*{Inverse parameterization}

An alternative parameterization of a multivariate Gaussian is in terms of its \href{https://en.wikipedia.org/wiki/Precision_(statistics)}{\emph{precision matrix}}, $\Lambda = \VV[\bX]^{-1}$. Similarly to above, we'll write $\lambda$ for the lower Cholesky factor of $\Lambda$, so $\lambda \lambda^t = \Lambda$. The parameterization is then specified by
\[
\bZ = \lambda\,(\bX - \mu)\ .
\]

In this parameterization, $\lambda$ is an \emph{inverse scale} or \emph{rate} parameter.
Solving for $z$ is of course now very simple, and the density becomes
\[
p_\bX(x) = |\lambda|\ p_\bZ(z) = |\lambda|\ p_\bZ(\lambda (x - \mu))\ .
\]

In exchange, forward sampling becomes awkward,
\[
\bX = \lambda^{-1}\bZ + \mu\ .
\]

\subsection*{Generalization}

Given an injective map $f\colon z \mapsto x$ and measures $\mu \ll_z \alpha$, the pushforward has density
\[
\frac{\dd\ f_*\mu}{\dd\ f_*\alpha}(x) 
= \frac{\dd \mu}{\dd \alpha}\left(f^{-1}(x)\right) 
= \frac{\dd \mu}{\dd \alpha}(z)\ .
\]

Note that there's no Jacobian to be found! This is because the measure and base measure are either both transformed, or both not. In fact, the Jacobian $\left|\frac{\dd x}{\dd z}\right|$ only comes into play when we compute ``across the transform'', and even then it's not in every case.

Changing our notation slightly, the general case is
\[
\frac{f_*\mu(\dd x)}{\alpha(\dd z)} 
= \frac{1}{
    \left|
        \frac{
            f_*\alpha(\dd x)
        }{
            \alpha(\dd z)
        }
    \right|
}
\frac{\mu(\dd z)}{\alpha(\dd z)}\ .
\]

This decomposes the problem into two subproblems. First we must compute 
$\left|
    \frac{
        f_*\alpha(\dd x)
    }{
        \alpha(\dd z)
    }
\right|$.
This plays the role of the determinant of the Jacobian, but is specific to the base measure $\alpha$. In particular, $\alpha$ might be a discrete measure, in which case this factor is one. Finally, we compute $\frac{\mu(\dd z)}{\alpha(\dd z)}$, which is just the (pre-transformation) density, more familiar from previous discussion as $\frac{\dd \mu}{\dd \alpha}(z)$.

For the Lebesgue case, if the Jacobian $\left|\frac{\dd x}{\dd z}\right|$ is not square but, say, $n\times k$ for $n>k$, the resulting measure will be \href{https://en.wikipedia.org/wiki/Embedding}{\emph{embedded}} into a $k$-dimensional affine subspace of $\RR^n$. This can be convenient for low-rank modeling, which can be important for high-dimensional data. If $\sigma$ has QR decomposition $\sigma=Q R$, we can use $|R|$ in place of $|\sigma|$ or $\left|\frac{\dd x}{\dd z}\right|$ above, since columns of $Q$ are orthonormal (so it's a change of basis and does not ``stretch'' the space). Our implementation is a variation of this that's equivalent but more efficient to compute.

% \moritz{We have to connect affine transforms with kernels I think. If we have infrastructure for affine functions we can use it to represent affine transformations of parameters and affine transformations of values. In particular, the parameter of a kernel is the value of the marginal measure in the marginal-conditional-joint decomposition} 

% \[
% \cC \sC \uC \fC \CC % calligraphic, script, upright, fraktur, blackboard-bold
% \]

%%%%%%%%%%%%%%%%%%%%%%%%%%%%%%%%%%%%%%%%%%%%%%%%
\section{Extensions}

Despite it being a very new package, \verb|MeasureTheory.jl| there is already active work to build up on it and to extend it.

\verb|PointProcesses.jl|\cite{Dalle_PointProcesses.jl} defines \href{https://en.wikipedia.org/wiki/Point_process}{\emph{point processes}}, in particular requiring the concept of \href{https://en.wikipedia.org/wiki/Random_measure}{\emph{random measure}}.

\verb|ManifoldMeasures.jl|\cite{manifoldmeasures} implement measures on a \href{https://en.wikipedia.org/wiki/Manifold}{\emph{manifold}}, using \href{https://en.wikipedia.org/wiki/Hausdorff_measure}{\emph{Hausdorff measure}} as the base measure.

\verb|MultivariateMeasures.jl|\cite{multivariatemeasures} gives high-performance implementations of \verb|logdensity| for multivariate measures, using \verb|LoopVectorization.jl|\cite{loopvectorization}.

\verb|Soss.jl|\cite{scherrer2020soss} is a \href{https://en.wikipedia.org/wiki/Probabilistic_programming}{\emph{probabilistic programming language}} that has recently adopted \verb|MeasureTheory.jl| as a foundation. In particular, every Soss \verb|Model| is an instance of \verb|AbstractMeasure|, and has another Soss model as its base measure.

As mentioned in Section~\ref{superposition}, \verb|ZigZagBoomerang.jl|\cite{https://doi.org/10.5281/zenodo.3931118} allows sampling with a spike and slab prior for sparse Bayesian inference and makes use the freedom to choose appropriate reference measures.  These models can be expressed using Soss.

\verb|Mitosis.jl|\cite{arxiv2010.03509} uses \verb|MeasureTheory.jl| to represent Bayesian networks via Markov kernels and defines transformations on those.

%%%%%%%%%%%%%%%%%%%%%%%%%%%%%%%%%%%%%%%%%%%%%%%%
\section{Related work}

While the vast majority of research in computational statistics works explicitly in terms of probability distributions, a few authors have addressed measures more generally:

\begin{itemize}
\item[$\bullet\!$]Borgstr\"om et al \cite{Borgstr_m_2013} describe the \verb|Fun| system in F\# in terms of \emph{measure transformer semantics}, but discusses only \emph{finite} measures.

\item[$\bullet\!$]Narayanan et al \cite{narayanan2016probabilistic} describe \verb|Hakaru|, a system for Bayesian modeling using measures. Here, a measure is a functional
\[
\mu[f] = \int f\ \dd\mu\ ,
\]
represented as a program. Hakaru's combinators are then expressed as compilers taking programs as the inputs.

\item[$\bullet\!$]Radul and Alexeev \cite{Radul2020} describe the \emph{base measure problem} of losing track of a base measure when applying a transformation, and suggest standardizing around Hausdorff measure as a solution. This problem doesn't arise for us, because the base measure is always taken into account.

\end{itemize}

In Julia \cite{bezanson2017julia}, the \verb|Distributions.jl| package \cite{Distributions.jl-2019} is very popular for computations on distributions. The drawbacks of \verb|Distributions.jl| are essentially those described in the first few sections of this paper. 
Current advantages over \verb|MeasureTheory| are the extensive range of distributions it implements and its popularity and familiarity to many Julia users. \verb|MeasureTheory.jl| currently has \verb|Distributions.jl| as a dependency, and uses it as a fall-back for many computations. For convenience, we also re-export the \verb|Distributions| module under the handle \verb|Dists|.

% Base \verb|R| does not treat distributions or random variables as first class objects (Strachey), for example a normal distribution is entirely handled via a set of functions \verb|rnorm|, \verb|dnorm|, \verb|pnorm| which are only associated by a naming convention, beyond acrobatic involving \verb|eval|
% \begin{lstlisting}
% > distname="norm"
% > eval(str2expression(paste("p",distname,sep="")))(1.645)
% [1] 0.9500151
% \end{lstlisting}

% The \verb|R| package \cite{distr-package} does implement distributions as ``subclasses of either \verb|AbscontDistribution| or \verb|DiscreteDistribution|, which themselves are again subclasses of \verb|Distribution|''. Beyond that, \verb|UnivarMixingDistribution| allows for mixtures between purely discrete and purely absolute continuous distribution without attempting to assign a density.
% \verb|UnivarLebDecDistribution|, a subclass of the former, uses the Lebesgue decomposition theorem to represent general univariate distributions first-class as mixture of absolute continuous and discrete part.\footnote{thus not allowing for distributions with singular continuous components such as the Cantor distribution, a very mild restriction.}

% This dichotomy between either discrete (supported on countable sets)  or absolute continuous breaks down in higher dimension.

\section{Conclusion}

We have introduced the concepts and implementation of \verb|MeasureTheory.jl|. This package is very new, so we expect there will be some changes as it matures. For this reason we have limited our discussion to aspects of the implementation we believe are relatively stable.

We hope this work can become a common foundation for probabilistic modeling in Julia. In particular, we believe this approach is especially well-suited for use in probabilistic programming, for which Julia has such a robust and active community.

We welcome discussion and community involvement with this package, as well as additional extensions to those we have described.

\input{bib.tex}

\end{document}

%% file: orcid.tex
\usepackage{tikz,xcolor,hyperref}
\definecolor{lime}{HTML}{A6CE39}
\DeclareRobustCommand{\orcidicon}{
	\begin{tikzpicture}
	\draw[lime, fill=lime] (0,0) 
	circle [radius=0.16] 
	node[white] {{\fontfamily{qag}\selectfont \tiny ID}};
	\draw[white, fill=white] (-0.0625,0.095) 
	circle [radius=0.007];
	\end{tikzpicture}
	\hspace{-2mm}
}
\foreach \x in {A, ..., Z}{\expandafter\xdef\csname orcid\x\endcsname{\noexpand\href{https://orcid.org/\csname orcidauthor\x\endcsname}
			{\noexpand\orcidicon}}
}

%% file: header.tex
\title{Applied Measure Theory for Probabilistic Modeling}

\author[1]{Chad Scherrer}
\author[2]{Moritz Schauer}
\affil[1]{Informative Prior}
\affil[2]{Chalmers and Gothenburg University, Sweden}

\keywords{Julia, Measure theory, Probability, Statistics}

\hypersetup{
pdftitle = {Applied Measure Theory for Probabilistic Modeling},
pdfsubject = {JuliaCon 2019 Proceedings},
pdfauthor = {Chad Scherrer, Moritz Schauer},
pdfkeywords = {Julia, Measure theory, Probability, Statistics},
}

%% file: bib.tex
\bibliographystyle{juliacon}
\bibliography{ref.bib}